# Observation of non-Hermitian boundary induced hybrid skin-topological effect excited by synthetic complex frequencies

Tianshu Jiang[1†*], Chenyu Zhang[1†], Ruo-Yang Zhang[2†], Yingjuan Yu[1], Zhenfu Guan[1], Zeyong Wei[1], Zhanshan Wang[1], Xinbin Cheng[1*], C. T. Chan[2*]

[1]*MOE Key Laboratory of Advanced Micro-Structured Materials, Shanghai Frontiers Science Center of Digital Optics, Institute of Precision Optical Engineering, and School of Physics Science and Engineering, Tongji University, Shanghai 200092, China*

[2]*Department of Physics, The Hong Kong University of Science and Technology, Hong Kong, China*

[†]These authors contributed equally to this work.

*Correspondence to: tsjiang@tongji.edu.cn; chengxb@tongji.edu.cn; phchan@ust.hk;

## Abstract

The hybrid skin-topological effect (HSTE) has recently been proposed as a mechanism where topological edge states collapse into corner states under the influence of the non-Hermitian skin effect (NHSE). However, directly observing this effect is challenging due to the complex frequencies of eigenmodes. In this study, we experimentally observe HSTE corner states using synthetic complex frequency excitations in a transmission line network. We demonstrate that HSTE induces asymmetric transmission along a specific direction within the topological band gap. Besides HSTE, we identify corner states originating from non-chiral edge states, which are caused by the unbalanced effective onsite energy shifts at the boundaries of the network. Furthermore, our results suggest that whether the bulk interior is Hermitian or non-Hermitian is not a key factor for HSTE. Instead, the HSTE states can be realized and relocated simply by adjusting the non-Hermitian distribution at the boundaries. Our research has deepened the understanding of a range of issues regarding HSTE, paving the way for advancements in the design of non-Hermitian topological devices.

**Introduction**

Over the past few decades, the exploration of Hermitian topological systems has marked a significant advancement in condensed matter physics[1-4] and photonics[5-9]. In particular, the chiral edge states in typical two-dimensional (2D) topological insulators give rise to unidirectional transport at the boundary. The study of these topological phases has underscored the elegant concept of bulk-boundary correspondence (BBC), which reveals the intrinsic connection between the existence of such edge states and non-zero bulk topological invariants, specifically Chern numbers.

Recently, there has been a burgeoning interest in open and nonconservative systems characterized by non-Hermitian Hamiltonians[10-15]. One of the most fascinating features in non-Hermitian systems is the non-Hermitian skin effect (NHSE)[16-33]. The NHSE is the phenomenon in which the bulk eigenstates collapse to the open boundaries forming skin modes. However, as bulk modes accumulate towards the boundary, the conventional Hermitian BBC is no longer applicable. To characterize the NHSE, the non-Bloch band theory[23,24] based on the general Brillouin zone (GBZ)[25-27] has been proposed. In contrast to the real-valued Bloch bands of Hermitian systems, the non-Hermitian spectra typically extend across the 2D complex energy plane resulting in the formation of point gaps. These point gaps have been demonstrated to be the topological origin of NHSE [21,25,30].

Furthermore, the interaction between chiral edge states and the non-Hermitian skin effect (NHSE) results in their hybridization, known as the hybrid skin-topological effect (HSTE)[34-47]. A distinctive characteristic of HSTE is the localization of topological chiral edge states induced by NHSE, while the bulk states remain extended. The occurrence of HSTE necessitates a point gap formed by edge states when imposing an open boundary condition (OBC) along either one dimension of a 2D lattice and maintaining a periodic boundary condition (PBC) along the other dimension. Subsequently, the edge states accumulate at the corner due to the NHSE.

HSTE was originally proposed theoretically using the 2D non-Hermitian Su-Schrieffer-Heeger (SSH) model[35]. It can also be implemented through coupled-wire construction[36], the non-Hermitian Haldane model[2,37,38], and the hyperbolic Haldane model[39]. Experimental observations of HSTE have been reported in circuit systems[40], photonic crystals[41,42], photonic crystals[43] and active matter systems[44], where both the edge and bulk exhibit non-Hermitian characteristics. While HSTE has been typically associated with non-Hermitian bulk properties, we will see that HSTE can be realized in Chern insulators solely through boundary modification while the bulk remains Hermitian. Additionally, the eigenfrequencies of HSTE are complex, presenting challenges for experimental implementation. Experimental observations of HSTE in the past required multiple real frequency excitation sources at different locations, followed by deducing the presence of HSTE using additional information, such as inequivalent responses[43]. However, these methods are indirect observations of HSTE and cannot measure the eigenstates of HSTE. Recent studies have demonstrated that complex frequency excitation can be synthesized from a series of real frequencies, which can compensate for losses in super-lenses and polariton propagation[48-50].

In this study, we experimentally observe HSTE in a transmission line network through the method of complex frequency excitation. Chiral edge states are achieved and can be described using the notion of angular momentum-orbital coupling in an effective Haldane model. By modulating unbalanced loss at the boundary, nontrivial point gaps emerge in the spectrum of edge modes for both armchair and zigzag edge strips within the topological bulk band gap regime. Consequently, HSTE is manifested in a finite-sized rectangular sample featuring both armchair and zigzag edges. The corner states of HSTE with complex eigenfrequencies are directly observed. In addition, we have also discovered non-chiral edge states in the real frequency range of the bulk band. Certain non-chiral edge states form distinct point gaps, leading to the NHSE.

Finally, we discuss the time-dependent characteristics of the complex frequency excitation method in detail in the Methods section.

## Results

**A conceptual tight-binding model.**

We first consider a conceptual tight-binding model exhibiting a particular form of angular-momentum-dependent topological transport[51] as shown in Fig. 1a. To be concise, we only show a hexagonal unit of honeycomb lattice that is periodic in the $x$-$y$ plane. The hexagon contains six inequivalent sublattices $A_1$, $A_2$, $A_3$, $B_1$, $B_2$ and $B_3$ (blue balls) and the connection between the nodes can be viewed as three layers of hexagons stacked on top of each other, where the subscripts 1, 2, 3 denote the layer number. The on-site energies of all the sublattices are set as zero. The gray and cyan lines denote the interlayer couplings $t_1$ and intralayer couplings $t_2$, respectively. The third layer is connected to the first layer so that the model is cyclic along the stacked direction. The sublattices $A_1$, $A_2$, $A_3$ are considered as internal constituents of the meta-atom $A$ and the $B_1$, $B_2$, $B_3$ compose the meta-atom $B$. Due to the 'head-to-tail' connection, the system can exhibit well-defined orbital angular momentum eigenmodes. In other words, the Hamiltonian can be blocked diagonalized into different sectors, each characterized by angular momentum eigenvalues $m = 0, \pm1$, in which the phase differences between the neighboring layers are $0, \pm 2\pi/3$, respectively. To realize angular-momentum-dependent topological transport, the couplings between the layers have specially designed spiral connections (see Fig. 1a) so that the node $A_i$ ($B_i$) is connected to other specific nodes $A_{(i+1) \bmod 3}$ ($B_{(i+1) \bmod 3}$) with $i = 1, 2, 3$. This twisted interconnectivity between the internal degrees of freedom of meta-atoms $A$ and $B$ is essential for the topological properties. Mathematically, this model can be viewed as an effective Haldane model with non-trivial gauge field[51] in sectors $m = \pm 1$ as shown in Fig. 1b. The eigenmodes of the model Hamiltonian has well-defined angular momentum quantum numbers and for sectors $m \neq 0$, the effective gauge field originates from the twisted interconnectivity between the internal degrees of freedom of meta-atoms $A$ and $B$. Figure 1b shows a 2D perspective, where the blue points denote the meta-atoms $A$ and $B$ and the solid black lines denote the couplings $t_2$ and the dashed gray arrows denote the effective non-reciprocal coupling $t_1 \exp(\pm i \cdot 2\pi/3)$ for angular momentum $m = \pm 1$, respectively. In this paper, we focus on the angular momentum sector $m = 1$. The Hamiltonian in $m = 1$ block can be written as

$$H(\mathbf{k}) = f_x(\mathbf{k})\sigma_x + f_y(\mathbf{k})\sigma_y + f_0(\mathbf{k})\cos(2\pi/3)\,\sigma_0 + f_z(\mathbf{k})\sin(2\pi/3)\,\sigma_z, \quad (1)$$

where $\mathbf{k} = (k_x, k_y)$ is the Bloch wavevector, $\sigma_0$ is the $2 \times 2$ identity matrix, $\sigma_x, \sigma_y$, and $\sigma_z$ are the Pauli matrices, and

$$f_0(\mathbf{k}) = t_1[4\cos\left(\tfrac{3}{2}k_x a\right)\cos\left(\tfrac{\sqrt{3}}{2}k_y a\right) + 2\cos(\sqrt{3}k_y a)],$$

$$f_x(\mathbf{k}) = t_2[1 + 2\cos\left(\tfrac{\sqrt{3}}{2}k_y a\right)\cos\left(\tfrac{3}{2}k_x a\right)],$$

$$f_y(\mathbf{k}) = 2t_2\cos\left(\tfrac{\sqrt{3}}{2}k_y a\right)\sin\left(\tfrac{3}{2}k_x a\right),$$

$$f_z(\mathbf{k}) = t_1[4\cos\left(\tfrac{3}{2}k_x a\right)\sin\left(\tfrac{\sqrt{3}}{2}k_y a\right) - 2\sin(\sqrt{3}k_y a)], \quad (2)$$

where $a$ represents the lattice constant, which is set to 1. Here $t_1$ and $t_2$ are both real numbers. In Eq. (1), the first two terms correspond to the Dirac cone induced by nearest neighbor hopping, which is commonly observed in hexagonal lattices. The third term corresponds to a global eigenvalue shift and the last term corresponds to the angular momentum-orbital couplings which open a non-trivial band gap.

Then we make the system non-Hermitian by introducing an on-site loss $-i\gamma$ on the sublattices $B_1$, $B_2$, $B_3$ (red balls) as shown in Fig. 1d. The corresponding effective 2D model becomes a non-Hermitian Haldane model with on-site loss $-i\gamma$ on meta-atom $B$ (red points) as shown in Fig. 1e. The corresponding effective Hamiltonian for the $m = 1$ block becomes

$$H_{\text{NH}}(\mathbf{k}) = H(\mathbf{k}) + \tfrac{1}{2}i\gamma(\sigma_z - \sigma_0), \quad (3)$$

It is known that the non-Hermitian Haldane model can exhibit the HSTE corner states[37,38,47], as will be discussed in the following sections.

**Transmission line network**

Now we realize the conceptual model using a transmission line network. Transmission lines are one dimensional co-axial cables and they can compose networks which have been used in the topological model study[51-53]. In a transmission line network, the voltage wave on each node satisfies the network equation[51],

$$-\psi_p \sum_q \coth(zl_{pq}) + \sum_q \frac{1}{\sinh(zl_{pq})} \psi_q = 0, \tag{4}$$

where $\psi_p$ and $\psi_q$ are wave functions at nodes $p$ and $q$, $l_{pq}$ is the cable length connecting node $p$ and node $q$, and $z = \left(\frac{i\omega}{c_0}\right)\sqrt{\varepsilon_r}$ with $\omega$, $c_0$ and $\varepsilon_r$ being the angular frequency, the speed of light in vacuum, and the relative permittivity of the dielectric medium in the coaxial cables, respectively. This equation is equivalent to a zero-energy nonlinear tight-binding equation where both the on-site terms $-\psi_p \sum_q \coth(zl_{pq})$ and hopping terms $\sum_q \frac{1}{\sinh(zl_{pq})} \psi_q$ are frequency dependent. The connecting nodes can be viewed as the atoms on sub-lattices and a transmission line corresponds to a coupling between two sites. The coupling coefficient depends on the frequency, length and permittivity of the cable.

In our simulations and experiments, the lengths of the inter-layer and intra-layer cables are 2.06 meters and 0.43 meters, respectively. The wave speed inside the cable is $0.66c_0$. Using these parameters and solving Eq. (4), we calculate the band structure of the transmission line network in the Hermitian model depicted in Fig. 1a. The resulting band structure is illustrated in Fig. 1c, where a topological band gap (gray zone) is observed between 31 MHz and 36.3 MHz.

To achieve the on-site loss in the non-Hermitian Haldane model (Fig. 1d), we connect terminators on sub-lattices $B_1$, $B_2$, $B_3$, respectively. The terminators create a leakage channel, resulting in the on-site loss. For the network with terminators, the network equation becomes (see Supplementary Note 1 and Supplementary Fig. 1 for more details),

$$-\psi_p [\sum_q \coth(zl_{pq}) - \delta] + \sum_q \frac{1}{\sinh(zl_{pq})} \psi_q = 0. \tag{5}$$

Here, $\delta$ represents the number of terminators on each lattice. For sub-lattices $A_1$, $A_2$ and $A_3$, $\delta = 0$; for sub-lattices $B_1$, $B_2$ and $B_3$, $\delta = 1$. In addition to the on-site loss at nodes, the coaxial cables also experience attenuation, causing the amplitude of the propagating waves to decay exponentially according to $\exp(-x/2L)$. Here, $x$ represents the propagation distance, and $L$ denotes the frequency-dependent absorption length (measured in meters), given by $L = \alpha f^{-\beta}$, where $\alpha = 338 \text{ m} \cdot \text{MHz}^\beta$, $\beta = 0.6123$, and $f$ is the wave frequency in MHz. The parameters of this empirical formula can be obtained through fitting to experimental measurements (see Supplementary Note 1 and Supplementary Fig. 2 for details). The corresponding non-Hermitian frequency dispersion is shown in Fig. 1f, which closely resembles that of the Hermitian Haldane model.

**HSTE in the transmission line network**

HSTE involves both the topological edge transportation and NHSE. To show these two effects simultaneously, we calculate the eigenfrequency spectra of the transmission line network supercells as shown in Figs. 2a-2e. As illustrated in the left panels of Figs. 2a-2c, the systems have a waveguide geometry and the supercells have periodic boundary conditions (PBCs) along the *x*-direction and open boundary conditions (OBCs) with an armchair configuration along the *y*-direction ("*x*-PBC/*y*-OBC"). To be concise, we present only the 2D honeycomb lattice, omitting details such as the inter-layer couplings in the figures. The red points denote on-site loss, implemented through terminators in the experiments. There are 9 rows of hexagons along the *y*-direction. The three supercells shown in Figs. 2a-2c corresponds to three different configurations, and they represent different loss distributions. The waveguide in Fig. 2a is formed by the unit cells depicted in Fig. 1e, featuring loss both within the bulk and at the boundary. We generate a 3D eigenvalue spectrum spanned by the Bloch wave vector $k_x$ and the complex frequency plane as shown in the middle panel of Fig. 2a. The eigenfrequencies of bulk states and edge states in the 3D parameter space are represented by gray and purple dots, respectively. Observing the projected band structure on the $k_x$- Re(*f*) plane (red dots in Fig. 2a), we observe two intersecting linear curves within the band gap, corresponding to the topological edge states at the upper and lower boundaries, respectively. The topological edge states are protected by the nontrivial Chern number of the original Hermitian Halmitonian[51]. Introducing loss throughout the entire sample (Fig. 2a) or on the edge (Fig. 2b) does not change the topological edge modes, as long as the topological line gap between two bands remains unclosed. It only introduces an imaginary part to their eigenfrequencies. The representative field patterns of the edge states are shown in the right panel of Figs. 2a, (i) and (ii). The yellow arrows denote the transport directions of the one-way going modes. Examining the projection onto complex frequency plane (with blue and pink dots representing bulk states and edge states respectively in Fig. 2a), we observe that the eigenfrequencies of the topological edge states at the upper and lower boundaries overlap along an arc inside the gap of the bulk continuum. The absence of a point gap within the bulk topological gap region suggests the absence of HSTE along the armchair boundaries when the *x*-periodic system is truncated into a finite lattice along the *y*-direction to form a waveguide. For the supercell shown in Fig. 2b, the loss is not introduced to the bulk but is confined solely to the boundary. Comparing the spectra in Figs. 2a and 2b, we can see significant differences in the complex frequencies of the bulk states, while the frequency dispersion of the edge states remains essentially the same. Neither of them exhibits a point gap within the topological gap. Therefore, the strip configuration shown in Fig. 2b also does not exhibit HSTE along the armchair boundaries.

The absence of nontrivial point gaps in the edge state spectra in Figs. 2a and 2b can be traced to the same symmetry origin: the loss distributions preserve the combined mirror-*y*, time-reversal, and Hermitian-conjugate symmetry, denoted as $M_y T^\dagger$. The $M_y T^\dagger$ symmetry of the two *x*-PBC/*y*-OBC strips gives rise to the spectral symmetry $f(k_x) = f(-k_x)$, which can rigorously connect the eigenfrequencies of the chiral edge states on the upper and lower boundaries inside the bulk band gap: $f_{\text{upper}}(k_x) = f_{\text{lower}}(-k_x)$. Consequently, the edge-state eigenfrequencies converge into a two-fold degenerate arc on the complex plane in each case. Besides the $M_y T^\dagger$ symmetry, the inversion symmetry $P$ can also lead to the spectral symmetry $f(k_x) = f(-k_x)$, thereby preventing edge states from forming nontrivial point gaps.

The two above examples demonstrate that the prerequisite for realizing a nontrivial point gap of edge states is to break all space-group symmetries that protect the spectral symmetry $f(k_x) = f(-k_x)$. We found that this can be achieved simply by adjusting the loss distribution on the lower boundary, thereby leading to the emergence of a nontrivial point gap for the topological edge states, as illustrated in the left panel of Fig. 2c. Specifically, a single lossy meta-atom is introduced on the lower boundary while the loss on the upper boundary remains unchanged. This imbalance in loss leads to a difference in the complex eigenfrequencies of the edge states localized on the upper and lower boundaries. The imaginary parts of the frequencies of the lower edge states have smaller amplitudes compared with cases of Fig. 2a, b while the ones of the upper edge states are unchanged. The upper and lower edge states together now form a point gap on the complex frequency plane as shown in the middle panel of Fig. 2c, indicating the existence of the NHSE along the loss unbalanced armchair boundaries. The winding of the point gap can be characterized by the winding number $\nu$:

$$\nu_\alpha = \oint_{BZ} \frac{dk_\alpha}{2\pi i} \frac{d}{dk_\alpha} \ln\left\{\prod_n [f_n(k_\alpha) - f_0]\right\}. \tag{6}$$

Here, $\alpha = x$ and $y$ indicates the direction along which the PBC is applied, the subscript $n$ is the index of the eigenfrequency, the multiplications are over all the eigenfrequencies $f_n(k_\alpha)$ of the strip at a fixed Bloch wave vector $k_\alpha$, and $f_0$ is the reference frequency within the point gap. In the spectrum shown in Fig. 2c, the cyan arrows indicate the winding direction as Bloch $k_x$ increases. According to Eq. (6), the winding number of the point gap within the bulk topological gap region is $\nu_x = 1$. Additionally, we notice that regardless of the presence/absence of a point gap, the chiral edge states in Figs. 2a-2c exhibit similar field patterns. However, the introduction of a point gap leads to differences in the imaginary parts of the eigenfrequencies of these edge states.

In the same way, we calculate the eigenfrequency spectra of the *x*-OBC/*y*-PBC supercells with zigzag boundary in the *y* direction, as shown in Figs. 2d-2e. There are 9 columns of hexagons along the *x*-directions. The onsite loss is imposed both in the bulk and boundaries in Fig. 2d while the loss is only imposed on the boundaries in Fig. 2e. Since the loss distributions in both cases break the symmetries that can protect the spectral symmetry $f(k_y) = f(-k_y)$, both their spectra exhibit point gaps with $\nu_y = 1$ within the bulk topological gap region, indicating the presence of the NHSE along the zigzag boundaries. Thus, we do not modify the loss on the boundaries. According to prior studies[41], a winding number $\nu_x = 1$ ($\nu_x = -1$) implies that the topological edge states of the *x*-PBC/*y*-OBC strip will collapse to the left (right) boundary when the strip is truncated into a finite-sized sample due to NHSE. Similarly, a winding number $\nu_y = 1$ ($\nu_y = -1$) indicates that the topological edge states of the *x*-OBC/*y*-PBC strip will collapse to the lower (upper) boundary. Under the combined effects of NHSE in both directions, HSTE corner states can form at a specific corner.

Combining the loss distributions of the structures in Figs. 2c and 2e, we design a 2D finite-sized transmission line network ("*x*-OBC/*y*-OBC") as shown in Fig. 3a. The size is 9 rows and 9 columns. The configuration shown in Fig. 3a can be interpreted as a topological 2D system with specialized boundary conditions imposed. The corresponding eigenfrequency spectrum in the complex plane is depicted by the blue points in Fig. 3b. For comparison, we also plotted the spectra of the *x*-PBC/*y*-OBC supercells of Fig. 2c (black solid points) and the *x*-OBC/*y*-PBC supercells of Fig. 2e (black dashed points) in the same figure. The eigenstates in the gray zone (the gap of the bulk continuum) indicate the topological edge states. The arrows on the boundaries in Fig. 3a indicate the transport direction of the topological edge states. According to the spectra in Figs. 2c and 2e, the edge states located on the left and upper boundaries are high-loss states (red arrows) while the edge states on the right and lower boundaries are low-loss states (blue arrows). If we apply a uniform gain to the entire sample, the entire complex spectrum will shift upwards along the Im(*f*) axis without altering its shape. This will not change the eigenstates of the sample. If we choose an appropriate gain value such that low-loss edge states become gain states, while high-loss edge states remain decay states, the wavefunction intensity of the gain edge states exponentially increases along the propagation direction, while the decay edge states exhibit exponential decay. Therefore, the wavefunction collapses to the lower left corner of the sample, forming HSTE corner states. Since adding overall gain does not alter the eigenstates, these HSTE corner states should also exist in the original sample, although they will decay as the whole system is dissipative. Figure 3c shows a representative field pattern of the HSTE corner states and Fig. 3d shows a field pattern of the bulk states. Based on the model in Fig. 3a, we further investigated networks with internal onsite losses (see Supplementary Fig. 3) and modified boundary loss configurations (see Supplementary Fig. 4). The results also demonstrate that only loss configurations at the network boundaries affect the existence of HSTE, while internal loss configurations have no effect. More examples of different boundary loss distributions, along with discussions regarding their symmetries, are presented in Supplementary Note 2 and Supplementary Figs. 5-6.

**The corner states from the non-chiral edge states**

In addition to the chiral edge states in the topological line gap, we notice the existence of non-chiral edge states in the real frequency range of the bulk band. The physical origin of the non-chiral edge states stems from the unbalanced effective onsite energies at the boundaries of the transmission line network. The detailed discussion can be seen in Supplementary Note 3 and Supplementary Figs. 7-9. In the spectrum of the armchair boundary strips in Fig. 2a, the frequencies of the non-chiral lower and upper edge states overlap in the lower left part of the complex frequency plane, forming two closed loops (and hence point gaps) respectively. As the Bloch wave vector $k_x$ increases from $-\pi/3$ to $\pi/3$, the eigen-frequencies of the lower (upper) edge states are observed to wind clockwise (counterclockwise) around the closed loop, indicating the localization of NHSE corner states at the lower right (upper left) corner if we have open boundary conditions in a *x*-OBC/*y*-OBC network. The representative field patterns of the lower and upper edge states are shown in (iii) and (iv) on the right panel of Fig. 2a. Note that these states are non-chiral, and hence can transmit both ways, as indicated by the yellow arrows.

In Fig. 2b, the onsite losses inside bulk are removed, but the properties of the edge states and corner states remain unaffected. The two overlapped closed loops persist. Moving to Fig. 2c, unbalanced losses lead to the separation of the closed loops. The imaginary parts of the frequencies of the lower edge states are raised, causing the corresponding loop to vanish, while the loop of the upper edge states remains unchanged. The winding number of the point gap formed by the upper edge states is $v_x = 1$. This indicates the presence of NHSE corner states at the upper left corner (Fig. 3e) and the absence of NHSE corner states at the lower right corner (Fig. 3f). For the zigzag boundary strips in Figs. 2d and 2e, the frequencies of the non-chiral edge states mix with the bulk states and fail to form closed loops. Consequently, NHSE corner states are absent for the non-chiral edge states along zigzag edges.

**Experimental demonstration with the synthesized complex frequency excitation**

Now we investigate the HSTE and corner states in the experiment. The experimental network photo is shown in Fig. 4a. The black lines denote the honeycomb lattice skeleton. The blue and red points mark the lattices without and with loss respectively. This network is a downscaled version of the model depicted in Fig. 3a, measuring $4 \times 5$ in size. The skeleton is shown in Fig. 4b and the corresponding complex eigenfrequency spectrum is shown in Fig. 4c.

First, we measure the field patterns of the eigenstates with complex eigen-frequencies. According to the linear response theory, the excitation efficiency of an eigenstate is proportional to $1/|\omega_0 - \omega_n|$ with $\omega_0$ and $\omega_n$ denoting the excitation frequency and the eigen-frequency concerned, respectively. When the eigen-frequency

$\omega_n$ has a large imaginary part, the eigenstate cannot be effectively excited at a real frequency even if $\omega_0 = \text{Re}[\omega_n]$ [54-57]. To solve this problem, we use a Fourier transformation method to synthesize a complex frequency exitation[48-50]. Considering the complex frequency excitation $J_0 e^{-i\omega_0 t}\theta(t)$ with angular frequency $\omega_0 = \omega_r - i\omega_i$ and a step function $\theta(t)$ ($\theta(t) = 1$ for $t \geq 0$, $\theta(t) = 0$ for $t < 0$), the time-domain response function $\psi(\omega_0, t)$ can be expanded into the integral of the real frequency-domain responses $A(\omega')$,

$$\psi(\omega_0, t) = \frac{1}{2\pi} \int_{-\infty}^{+\infty} \frac{iA(\omega')}{\omega' - \omega_0} e^{-i\omega' t} d\omega'. \tag{7}$$

In practice, it can be approximately approached by the discrete form:

$$\psi(\omega_0, t) \approx \sum_j \frac{iA(\omega_j)}{\omega_j - \omega_0} e^{-i\omega_j t} \Delta\omega/2\pi. \tag{8}$$

Here, the summation is taken over a finite spectrum range and the response given by Eq. (8) exhibits an overall temporal periodicity of $2\pi/\Delta\omega$. In experiments, we measured frequencies from 10 MHz to 50 MHz in 0.05 MHz intervals. More details about the complex frequency synthesis can be seen in Methods and Figs. 5-7.

The complex frequency excited HSTE corner states are shown in Figs. 4d, in which the excitation complex frequency is set as $f = (33.49 - 1.19i)$ MHz. The upper and lower panels show the experimental and simulated results respectively, which agree with each other very well. Upon observing the field patterns, it can be seen that the HSTE corner state is excited, where the waves are primarily localized at the lower left corner. Due to the limited size of the experimental network, the interaction between the opposite boundaries leads to a noticeable strength in the wavefunction at the upper left corner as well (see Supplementary Note 4 and Supplementary Figs. 10-12 for more details). Apart from the HSTE corner states, the upper left corner states originating from the non-chiral edge state on the upper boundary is also observed with frequency $f = (27.34 - 1.78i)$ MHz as shown in Figs. 4e. In Fig. 4f, we also demonstrate that the edge state with frequency $f = (27.72 - 0.94i)$ MHz on the lower boundary is not confined at a conner due to the absence of a point gap at this frequency as shown in Fig. 2c.

Although the HSTE corner states can be observed through complex frequency synthesis with high efficiency, we still expect observable phenomenon in the real frequency regime although the excitation efficiency is expected to be lower. One such phenomenon is the asymmetric transmission along the anti-diagonal direction. The schematics for measuring the transmission are shown in the inset of Fig. 4g. The $S_{21}$ denotes the transmission from the upper right corner to the lower left corner while the $S_{12}$ denotes the transmission along the opposite direction. Since the HSTE corner states only exist at the lower left corner, using this corner as an output point results in a significantly higher transmission rate compared to using it as an input point. It causes asymmetric transmission in the topological band gap. The experimental measured transmission spectra are shown in Fig. 4g. The asymmetric transmission is verified that the $|S_{21}|$ and $|S_{12}|$ are different in the topological band gap.

## Discussion

In this work, we conducted experiments using a transmission line network to observe HSTE corner states and non-chiral edge/corner states. For the first time, we introduced a synthetic complex frequency approach to directly observe the eigenstates with complex eigen-frequencies in non-Hermitian systems. Additionally, we observed asymmetric transmission along a specific direction in the band gap as a manifestation of HSTE in the real frequency domain. Furthermore, our results demonstrate that a non-Hermitian bulk is not required for HSTE. Simply by adjusting the non-Hermitian boundary configuration, we cannot only attain the HSTE states but also manipulate their localization positions (see Supplementary Note 5 and Supplementary Fig. 13 for more details). The HSTE states have been confirmed to be robust against disorder (see Supplementary Note 6 and Supplementary Fig. 14 for more details). This insight is significant for applications such as topological laser design, suggesting that bulk gain materials within the system can be omitted, leading to substantial simplification in system design and realization. Furthermore, the tunability of the localized positions of HSTE states provides additional flexibility in the design of functional non-Hermitian devices, such as position sensors and wave shapers. In addition, we explore the effects of sample shapes (see Supplementary Note 7 and Supplementary Figs. 15-17), angular momenta (see Supplementary Note 8 and Supplementary Figs. 18-19), and larger winding numbers on HSTE (see Supplementary Note 9 and Supplementary Fig. 20) in the Supplementary Information.

## Methods

**The physical meaning and derivation of complex frequency synthesis methods**

In the main text, we use the Fourier transformation to synthesize the complex frequency. In this section, we will discuss the physical meaning of this method and its mathematical derivation in detail.

In non-Hermitian systems with loss, the eigenfrequencies are typically complex numbers, where the imaginary part of an eigenfrequency is related to the inverse of the lifetime of the damped eigenstate: $\text{Im}(\omega_s) = 1/\tau_s$. To counteract the loss and maintain a steady state, one method is to introduce an external driving force, such as gain materials. Another approach is to use complex frequency excitation to match the complex eigenfrequency. However, a genuine complex frequency itself is not physically realizable as the amplitude will exponentially approach infinity as $t \to -\infty$. Complex frequency excitation is a mathematical notion that can be used to recover the information lost due to non-Hermitian loss, rather than a physical entity per se.

In some prior studies, complex frequency excitation is achieved through temporal attenuation excitation[56,57], where the imaginary part of the frequency corresponds to temporal decay. This method requires time-domain measurements of the responses, which can be challenging in high-frequency systems. Recently, a new method of complex frequency excitation based on multiple real frequency excitations has been proposed[48-50]. In this approach, complex frequency excitation is synthesized using Fourier transformation between the frequency domain and the time domain. Measurements are conducted under static wave excitations in the frequency domain, which is easier to implement experimentally. By using Fourier transformation, we can convert the responses from the frequency domain to the time domain, thereby avoiding the direct measurements of transient states.

For completeness, we provide a detailed illustration of the mathematics behind this idea. Considering a source $J(x',t)$ with complex frequency $\omega_0 = \omega_r - i\omega_i$, it can be expressed as:

$$J(x',t) = J_0(x')e^{-i\omega_0 t}\theta(t). \tag{9}$$

Here $\theta(t)$ is a step function with $\theta(t) = 1$ for $t \geq 0$ and $\theta(t) = 0$ for $t < 0$, which expresses the fact that any excitation has a starting time ($t = 0$).

The source can be expressed in the frequency domain through the Fourier transform:

$$\tilde{J}(x',\omega') = \int_{-\infty}^{\infty} J_0(x')e^{-i\omega_0 t}\theta(t)e^{i\omega' t}dt = J_0(x')\frac{i}{\omega'-\omega_0}. \tag{10}$$

The corresponding response function in the time domain can be expressed as:

$$\psi(x,\omega_0,t) = \frac{1}{2\pi}\int_{-\infty}^{\infty}d\omega'\int dx'\,\tilde{J}(x',\omega')\tilde{G}(x,x',\omega')e^{-i\omega' t}. \tag{11}$$

Here, $\tilde{G}(x,x',\omega')$ is the Green's function in the frequency domain. Substituting Eq. (10) into Eq. (11), the response function becomes:

$$\begin{aligned}\psi(x,\omega_0,t) &= \frac{1}{2\pi}\int_{-\infty}^{\infty}d\omega'\int dx'\,\frac{i}{\omega'-\omega_0}J_0(x')\tilde{G}(x,x',\omega')e^{-i\omega' t}\\ &= \frac{1}{2\pi}\int_{-\infty}^{\infty}d\omega'\,\frac{i}{\omega'-\omega_0}A(x,\omega')e^{-i\omega' t}.\end{aligned} \tag{12}$$

Here, $A(x,\omega') = \int dx' J_0(x')\tilde{G}(x,x',\omega')$ is the real frequency-domain response at position $x$. In practice, Eq. (12) is converted to the discrete form:

$$\psi(\omega_0,t) \approx \sum_{j=1}^{N}\frac{iA(\omega_j)}{\omega_j-\omega_0}e^{-i\omega_j t}\Delta\omega/2\pi. \tag{13}$$

Here, $x$ is omitted and the summation range is from $\omega_1$ to $\omega_N$. Due to the discretization, the strength of the response function exhibits periodicity with a period of $T = 2\pi/\Delta\omega$. The proof is as follows:

$$\psi(\omega_0, t+T) = \sum_{j=1}^{N} \frac{iA(\omega_j)}{\omega_j - \omega_0} e^{-i\omega_j(t+T)} \Delta\omega/2\pi$$

$$= \sum_{j=1}^{N} \frac{iA(\omega_j)}{\omega_j - \omega_0} \left(e^{-i\omega_j t} \cdot e^{-i[\omega_1 + (j-1)\Delta\omega]T}\right) \Delta\omega/2\pi$$

$$= e^{-i\omega_1 T} \cdot \sum_{j=1}^{N} \frac{iA(\omega_j)}{\omega_j - \omega_0} e^{-i\omega_j t} \Delta\omega/2\pi$$

$$= e^{-i\omega_1 T} \cdot \psi(\omega_0, t) \tag{14}$$

To show the periodicity more clearly, we plot the temporal evolutions of the responses at specific positions for the experimental configuration. The signal of $f = (33.49 - 1.19i)$ MHz is input from the position indicated by the gray circle as shown in Fig. 5a. The time-dependent wave intensities obtained from Eq. (13) at the positions indicated by the cyan and pink circles are shown in Figs. 5b and 5c, respectively. The upper and lower panels are experimental and simulated results, respectively. It is obvious that the response intensities have a period of $T = 20\ \mu s$, which is consistent with the frequency interval 0.05 MHz. As we will see below, the eigenstates excited by the corresponding complex frequencies can only be observed within specific time windows, which approximately coincide with the peaks of the response functions shown in Figs. 5b and 5c. Therefore, the selected frequency intervals $\Delta\omega$ should ensure that the peaks of the response functions separated by one period $T = 2\pi/\Delta\omega$ do not overlap.

Next, we show how to process the experimental data using Eq. (13) and plot the temporal evolution of the strength field patterns. We selected different excitation frequencies and source positions, as shown in Figs. 6a-6d. In Fig. 6a, the excitation frequency is chosen at the eigenfrequency of a corner state, $f = (33.49 - 1.19i)$ MHz. The excitation positions are located at the lower right corners. Analyzing the evolution process, we noted that once the signal entered the transmission line network, it followed a clockwise propagation along the boundary. Upon reaching the lower left corner, it remained in a HSTE corner state for a considerable duration before dissipating due to losses. In the main text, we choose $t = 0.57$ $\mu s$ for Fig. 4d. To correspond with Figs. 4e and 4f in the main text, we also plot the time-dependent field patterns for the exitation frequencies $f = (27.34 - 1.78i)$ MHz and $f = (27.72 - 0.94i)$ MHz, as shown in Figs. 6b and 6c, respectively. In the main text, we choose the transient field profiles at $t = 0.24$ and $0.39$ $\mu s$ in Figs. 6b and 6c to be presented in Figs. 4e and 4f, respectively. For comparison, Fig. 6d illustrates the excitation result at a real frequency $f = 33.49$ MHz, which corresponds to the real part of the eigenfrequency $(33.49 - 1.19i)$ MHz of the same corner state excited in Fig. 6a. Consequently, it is clear that the excitation field fails to effectively exhibit a HSTE corner state pattern, in stark contrast to the result depicted in Fig. 6a. This demonstrates both the efficacy and necessity of employing synthetic complex frequencies to excite eigenstates characterized by complex eigenfrequencies.

**The temporal evolution of the response function**

From the previous discussion, we know that the response function obtained from Eq. (12) is time dependent. In this section, we will provide a detailed explanation of the evolution of this response function over time and discuss why this function allows us to observe the eigenstates corresponding to complex frequencies within a certain time window.

According to the Residue Theorem, the response function $\psi(x, \omega_0, t)$ in Eq. (12) can be expressed as:

$$\psi(x, \omega_0, t) = -2\pi i [\text{Res}(F(x, \omega'), \omega_0) + \sum_s \text{Res}(F(x, \omega'), \omega_s)]. \quad (15)$$

Here, $F(x, \omega') = iA(x, \omega')e^{-i\omega' t}/2\pi(\omega' - \omega_0)$, which has poles located both at the excitation frequency $\omega_0$ and at the eigenfrequencies $\omega_s$ of the system. The reason for the negative sign in this equation is that all poles are located in the lower half of the complex frequency plane, so the integration path is in the clockwise direction (as shown in Fig. 7a). To approximate the contour integral by the integral over an interval on the real axis, the integral over the arc segment (yellow curve) must be close to zero, which requires that the arc is as far away from the poles as possible. This means that the upper and lower limits of the discretized integration in Eq. (13) should be as far away from the eigenfrequencies as possible.

Using the Green's function of the tight-binding model, we can express the real frequency-domain responses $A(x, \omega')$ as:

$$A(x, \omega') = \int dx' J_0(x') \tilde{G}(x, x', \omega') = \int dx' J_0(x') \sum_s \frac{|\psi_s^R(x)\rangle\langle\psi_s^L(x)|}{\omega' - \omega_s}. \quad (16)$$

Thus, the integrand $F(x, \omega')$ is expressed as:

$$F(x, \omega') = \frac{iA(x, \omega')e^{-i\omega't}}{2\pi(\omega' - \omega_0)} = \sum_s \frac{iB_s(x)e^{-i\omega't}}{2\pi(\omega' - \omega_0)(\omega' - \omega_s)}. \quad (17)$$

Here, $B_s(x) = \int dx' J_0(x') |\psi_s^R(x)\rangle\langle\psi_s^L(x)|$ represents the excited field pattern associated with the eigenfrequency $\omega_s$. Therefore, it corresponds to the eigenstate field pattern for the eigenfrequency $\omega_s$.

First, we assume that $\omega_0$ is unequal to any of the eigenfrequencies $\omega_s$ so that all the poles are of first-order. The response function $\psi(x, \omega_0, t)$ is:

$$\psi(x, \omega_0, t) = -2\pi i \left[ \text{Res}(F(x, \omega'), \omega_0) + \sum_s \text{Res}(F(x, \omega'), \omega_s) \right]$$

$$= \sum_s \frac{B_s(x)e^{-i\omega_0 t}}{\omega_0 - \omega_s} + \sum_s \frac{B_s(x)e^{-i\omega_s t}}{\omega_s - \omega_0}$$

$$= \sum_s \frac{(e^{-i\omega_s t} - e^{-i\omega_0 t})}{\omega_s - \omega_0} B_s(x). \quad (18)$$

Specifically, if the excitation frequency $\omega_0$ approaches one of the eigenfrequencies $\omega_{s_0}$, the response will converge to the form:

$$\lim_{\omega_0 \to \omega_{s_0}} \psi(x, \omega_0, t) = \lim_{\omega_0 \to \omega_{s_0}} \frac{(e^{-i\omega_{s_0} t} - e^{-i\omega_0 t})}{\omega_{s_0} - \omega_0} B_{s_0}(x) + \sum_{s \neq s_0} \frac{(e^{-i\omega_{s_0} t} - e^{-i\omega_0 t})}{\omega_{s_0} - \omega_0} B_s(x)$$

$$= -ite^{-i\omega_{s_0} t} B_{s_0}(x) + \sum_{s \neq s_0} \frac{(e^{-i\omega_s t} - e^{-i\omega_{s_0} t})}{\omega_s - \omega_{s_0}} B_s(x). \quad (19)$$

The first term represents the response of the eigenstate of the excitation frequency $\omega_{s_0}$, while the second term represents the responses of the eigenstates of other eigenfrequencies $\omega_s$. Their coefficients evolve over time with coefficients of $C_{s_0} = -ite^{-i\omega_{s_0}t}$ and $C_s = (e^{-i\omega_s t} - e^{-i\omega_{s_0}t})/(\omega_s - \omega_{s_0})$, respectively.

We then calculated the absolute values of the coefficients over time, as illustrated in Fig. 7b. This calculation employs the tight-binding model, configured identically to that depicted in Fig. 4b of the main text. The parameters are set as $t_1 = -1$, $t_2 = -2$ and $\gamma = 1$. The corresponding energy spectrum is presented in the subplot of Fig. 7b. To ensure consistency with earlier statements, we will refer to energy as frequency in the latter part of this section. The excitation frequency $\omega_{s_0}$ is marked by a circle, and the color of each eigenfrequency point represents its distance $d$ from the excitation frequency $\omega_{s_0}$ on the complex frequency plane. We calculated the time-dependent curves of the coefficients $|C_{s_0}|$ and $|C_s|$ for all the eigenfrequencies as shown in Fig. 7b. The colors of these curves correspond one-to-one with the colors of the eigenfrequencies. The black dashed curve represents the coefficient $|C_{s_0}|$, which is significantly larger than the other coefficients $|C_s|$ over an extended period (about $4 < t < 10$), dominating the response. This indicates that within this time window, we can clearly observe the eigenstate of $\omega_{s_0}$. It explains why we can use the synthetic complex frequency excitation method to observe the eigenstates of complex frequencies.

## Data availability

The data that support the findings of this study are available from the corresponding authors on request.

## Reference

1. Halperin, B. I. Quantized Hall conductance, current-carrying edge states, and the existence of extended states in a two-dimensional disordered potential. *Phys. Rev. B* **25**, 2185–2190 (1982).20

2. Haldane, F. D. M. Model for a Quantum Hall Effect without Landau Levels: Condensed-Matter Realization of the 'Parity Anomaly'. *Phys. Rev. Lett.* **61**, 2015–2018 (1988).

3. Hatsugai, Y. Chern number and edge states in the integer quantum Hall effect. *Phys. Rev. Lett.* **71**, 3697–3700 (1993).

4. Qi, X.-L. & Zhang, S.-C. Topological insulators and superconductors. *Rev. Mod. Phys.* **83**, 1057–1110 (2011).

5. Haldane, F. D. M. & Raghu, S. Possible realization of directional optical waveguides in photonic crystals with broken time-reversal symmetry. *Phys. Rev. Lett.* **100**, 013904 (2008).

6. Wang, Z., Chong, Y., Joannopoulos, J. D. & Soljačić, M. Observation of unidirectional backscattering-immune topological electromagnetic states. *Nature* **461**, 772-775 (2009).

7. Hafezi, M., Mittal, S., Fan, J., Migdall, A. & Taylor, J. M. Imaging topological edge states in silicon photonics. *Nat. Photon.* **7**, 1001-1005 (2013).

8. Khanikaev, A. B. *et al.* Photonic topological insulators. *Nat. mater.* **12**, 233-239 (2013).

9. Ozawa, T. *et al.* Topological photonics. *Rev. Mod. Phys.* **91**, 015006 (2019).

10. Bender, C. M. & Boettcher, S. Real spectra in non-Hermitian Hamiltonians having P T symmetry. *Phys. Rev. Lett.* **80**, 5243-5246 (1998).

11. Rüter, C. E. *et al.* Observation of parity–time symmetry in optics. *Nat. phys.* **6**, 192-195 (2010).

12. Feng, L., Wong, Z. J., Ma, R.-M., Wang, Y. & Zhang, X. Single-mode laser by parity-time symmetry breaking. *Science* **346**, 972-975 (2014).

13. Ashida, Y., Gong, Z. & Ueda, M. Non-Hermitian physics. *Adv. Phys.* **69**, 249-435 (2020).

14. Bergholtz, E. J., Budich, J. C. & Kunst, F. K. Exceptional topology of non-Hermitian systems. *Rev. Mod. Phys.* **93**, 015005 (2021).

15. Meng, H., Ang, Y. S. & Lee, C. H. Exceptional points in non-Hermitian systems: Applications and recent developments. *Appl. Phys. Lett.* **124,** 060502 (2024).

16. Yao, S. & Wang, Z. Edge states and topological invariants of non-Hermitian systems. *Phys. Rev. Lett.* **121**, 086803 (2018).

17. Yao, S., Song, F. & Wang, Z. Non-Hermitian Chern bands. *Phys. Rev. Lett.* **121**, 136802 (2018).

18. Lee, T. E. Anomalous Edge State in a Non-Hermitian Lattice. *Phys. Rev. Lett.* **116**, 133903 (2016).

19. Xiong, Y. Why does bulk boundary correspondence fail in some non-Hermitian topological models. *J. Phys. Commun.* **2**, 035043 (2018).

20. Helbig, T. *et al.* Generalized bulk–boundary correspondence in non-Hermitian topolectrical circuits. *Nat. Phys.* **16**, 747-750 (2020).

21. Okuma, N., Kawabata, K., Shiozaki, K. & Sato, M. Topological origin of non-Hermitian skin effects. *Phys. Rev. Lett.* **124**, 086801 (2020).

22. Zhang, X., Zhang, T., Lu, M.-H. & Chen, Y.-F. A review on non-Hermitian skin effect. *Adv. Phys. X* **7**, 2109431 (2022).

23. Kawabata, K., Okuma, N. & Sato, M. Non-Bloch band theory of non-Hermitian Hamiltonians in the symplectic class. *Phys. Rev. B* **101**, 195147 (2020).

24. Yokomizo, K. & Murakami, S. Non-Bloch bands in two-dimensional non-Hermitian systems. *Phys. Rev. B* **107**, 195112 (2023).

25. Zhang, K., Yang, Z. & Fang, C. Correspondence between Winding Numbers and Skin Modes in Non-Hermitian Systems. *Phys. Rev. Lett.* **125**, 126402 (2020).

26. Yang, Z., Zhang, K., Fang, C. & Hu, J. Non-Hermitian Bulk-Boundary Correspondence and Auxiliary Generalized Brillouin Zone Theory. *Phys. Rev. Lett.* **125**, 226402 (2020).

27. Yi, Y. & Yang, Z. Non-Hermitian Skin Modes Induced by On-Site Dissipations and Chiral Tunneling Effect. *Phys. Rev. Lett.* **125**, 186802 (2020).

28. Zirnstein, H.-G., Refael, G. & Rosenow, B. Bulk-Boundary Correspondence for Non-Hermitian Hamiltonians via Green Functions. *Phys. Rev. Lett.* **126**, 216407 (2021).

29. Zhang, K., Yang, Z. & Fang, C. Universal non-Hermitian skin effect in two and higher dimensions. *Nat. Commun.* **13**, 2496 (2022).

30. Ghatak, A. & Das, T. New topological invariants in non-Hermitian systems. *J. Phys. Condens. Matter* **31**, 263001 (2019).

31. Gao, H. *et al.* Anomalous Floquet non-Hermitian skin effect in a ring resonator lattice. *Phys. Rev. B* **106**, 134112 (2022).

32. Yin, C., Jiang, H., Li, L., Lü, R. & Chen, S. Geometrical meaning of winding number and its characterization of topological phases in one-dimensional chiral non-Hermitian systems. *Phys. Rev. A* **97**, 052115 (2018).

33. Lin, R., Tai, T., Li, L. & Lee, C. H. Topological non-Hermitian skin effect. *Front. Phys.* **18**, 53605 (2023).

34. Zhu, W. & Li, L. A brief review of hybrid skin-topological effect. *J. Phys. Condens. Matter* **36,** 253003 (2024).

35. Lee, C. H., Li, L. & Gong, J. Hybrid higher-order skin-topological modes in nonreciprocal systems. *Phys. Rev. Lett.* **123**, 016805 (2019).

36. Li, L., Lee, C. H. & Gong, J. Topological switch for non-Hermitian skin effect in cold-atom systems with loss. *Phys. Rev. Lett.* **124**, 250402 (2020).

37. Li, Y., Liang, C., Wang, C., Lu, C. & Liu, Y.-C. Gain-loss-induced hybrid skin-topological effect. *Phys. Rev. Lett.* **128**, 223903 (2022).

38. Zhu, W. & Gong, J. Hybrid skin-topological modes without asymmetric couplings. *Phys. Rev. B* **106**, 035425 (2022).

39. Sun, J., Li, C.-A., Feng, S. & Guo, H. Hybrid higher-order skin-topological effect in hyperbolic lattices. *Phys. Rev. B* **108**, 075122 (2023).

40. Zou, D. *et al.* Observation of hybrid higher-order skin-topological effect in non-Hermitian topolectrical circuits. *Nat. Commun.* **12**, 7201 (2021).

41. Liu, G.-G. *et al.* Localization of chiral edge states by the non-Hermitian skin effect. *Phys. Rev. Lett.* **132**, 113802 (2024).

42. Sun, Y. *et al.* Photonic Floquet skin-topological effect. *Phys. Rev. Lett.* **132**, 063804 (2024).

43. Wu, J. *et al.* Spin-Dependent Localization of Helical Edge States in a Non-Hermitian Phononic Crystal. *Phys. Rev. Lett.* **133**, 126601 (2024).

44. Palacios, L. S. *et al.* Guided accumulation of active particles by topological design of a second-order skin effect. *Nat. Commun.* **12**, 4691 (2021).

45. Li, Y., Lu, C., Zhang, S. & Liu, Y.-C. Loss-induced Floquet non-Hermitian skin effect. *Phys. Rev. B* **108**, L220301(2023).

46. Wang, L.-W., Lin, Z.-K. & Jiang, J.-H. Non-Hermitian topological phases and skin effects in kagome lattices. *Phys. Rev. B* **108**, 195126 (2023).

47. Ma, X.-R. *et al.* Non-Hermitian chiral skin effect. *Phys. Rev. Research* **6**, 013213 (2024).

48. Guan, F. *et al.* Overcoming losses in superlenses with synthetic waves of complex frequency. *Science* **381**, 766–771 (2023).

49. Guan, F. *et al.* Compensating losses in polariton propagation with synthesized complex frequency excitation. *Nat. Mater.* **23**, 506–511 (2024).

50. Zeng, K. *et al.* Synthesized complex-frequency excitation for ultrasensitive molecular sensing. *eLight* **4**, 1 (2024).

51. Jiang, T. *et al.* Experimental demonstration of angular momentum-dependent topological transport using a transmission line network. *Nat. Commun.* **10**, 434 (2019).

52. Guo, Q. *et al.* Experimental observation of non-Abelian topological charges and edge states. *Nature* **594**, 195-200 (2021).

53. Jiang, T. *et al.* Four-band non-Abelian topological insulator and its experimental realization. *Nat. Commun.* **12**, 6471 (2021).

54. Li, H., Mekawy, A., Krasnok, A. & Alù, A. Virtual parity-time symmetry. *Phys.*

*Rev. Lett.* **124**, 193901 (2020).

55. Schomerus, H. Fundamental constraints on the observability of non-Hermitian effects in passive systems. *Phys. Rev. A* **106**, 063509 (2022).

56. Gu, Z. *et al.* Transient non-Hermitian skin effect. *Nat. Commun.* **13**, 7668 (2022).

57. Kim, S., Peng, Y.-G., Yves, S. & Alù, A. Loss compensation and superresolution in metamaterials with excitations at complex frequencies. *Phys. Rev. X* **13,** 041024 (2023).

## Acknowledgment

The work is supported by the National Natural Science Foundation of China (12304345). The work done in Hong Kong is supported by Hong Kong RGC through grants 16307821 and AoE/P-502/20.

## Author contributions

T.J. conceived the idea, developed the theory, and conducted the experiments. C.Z. assisted with numerical simulations and the experiments. R.-Y.Z. provided important suggestions for the theory. T.J., C.Z., R.-Y.Z., and C.T.C. wrote the manuscript. Y.Y., Z.G., Z.Y.W. and Z.S.W. provided support. T.J., X.C., and C.T.C. supervised the project.

## Competing interests

The authors declare no competing interests.

# Figures

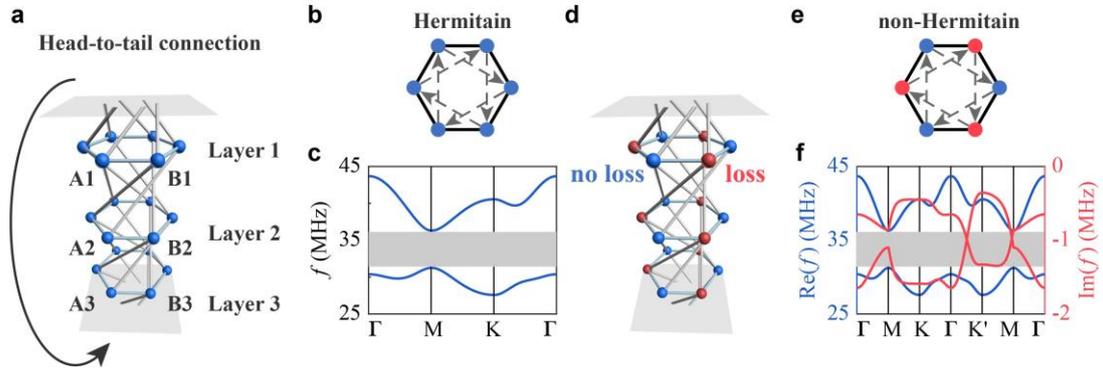

**Fig. 1| Models and band structures. a** The conceptual tight-binding model which exhibits angular momentum-orbital coupling. **b** The model in (**a**) can be viewed as an effective 2D Haldane model. **c** The frequency band structure of the model in (**a**) for the angular momentum sector $m = 1$. **d** Loss is introduced to the sublattices $B_1$, $B_2$, $B_3$ of the model in (**a**). **e** The model in (**d**) can be viewed as an effective 2D non-Hermitian Haldane model. **f** The frequency band structure of the model in (**d**) for the angular momentum sector $m = 1$.

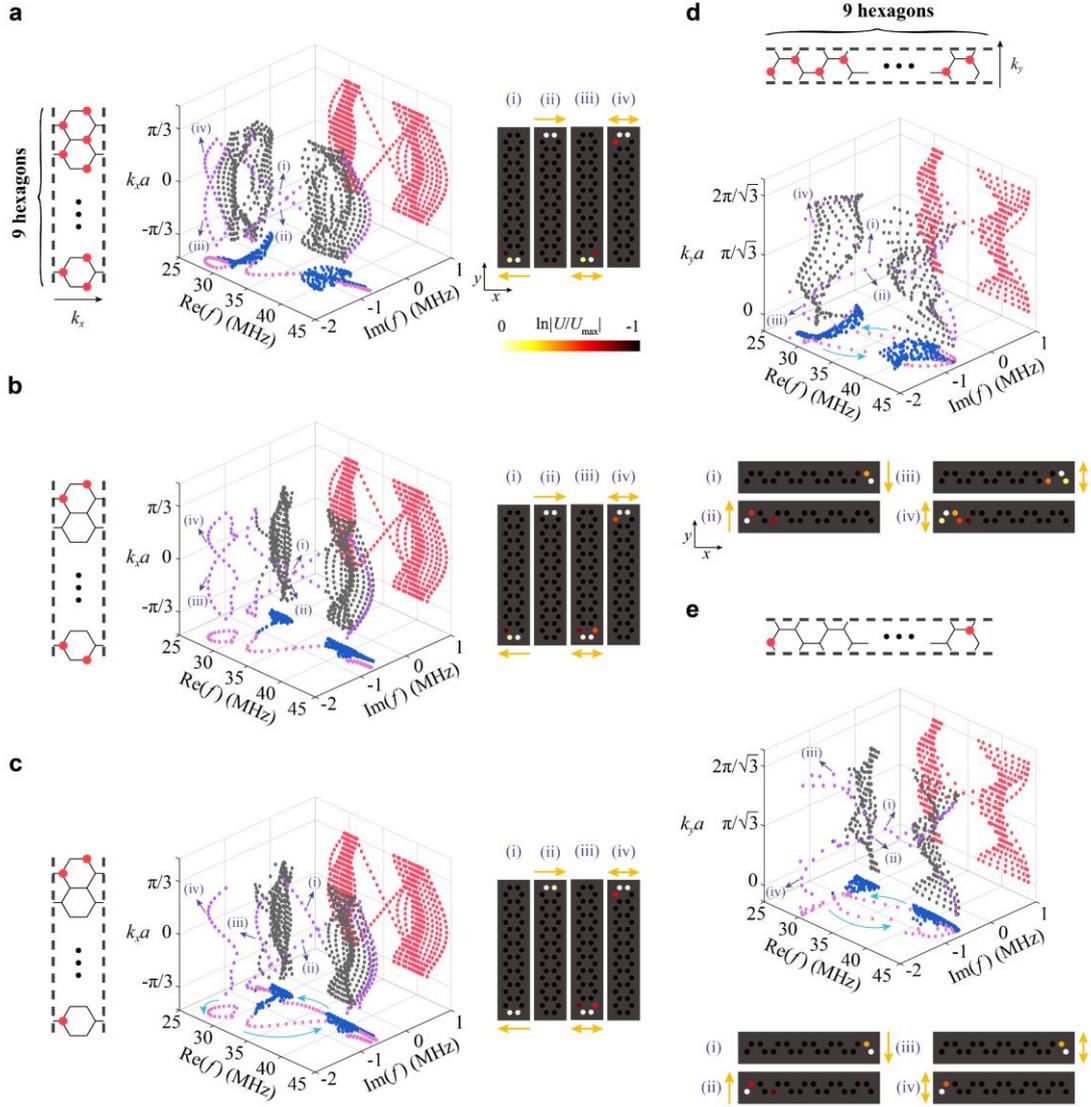

**Fig. 2| Eigenfrequencies and edge states of the supercells. a-c** The eigenfrequency spectra and the edge states of the *x*-PBC/*y*-OBC supercells with armchair boundary configurations. The left panels are the supercell configurations. The middle panels are the 3D eigenfrequency spectra. The right panels are the field patterns of select eigenstates. The color of each node denotes the value of $\ln|U/U_{max}|$, where $U$ is the voltage, and $U_{max}$ is the maximum voltage among all nodes. The Roman numerals correspond to specific frequency points in the 3D spectra. Patterns (i) and (ii) are chiral edge states, whereas patterns (iii) and (iv) are non-chiral edge states. **d-e** The eigenfrequency spectra and the edge states of the *x*-OBC/*y*-PBC supercells with zigzag boundary configurations. The upper, middle and lower panels are the supercell configurations, the 3D eigenfrequency spectra and selected eigenstates, respectively.

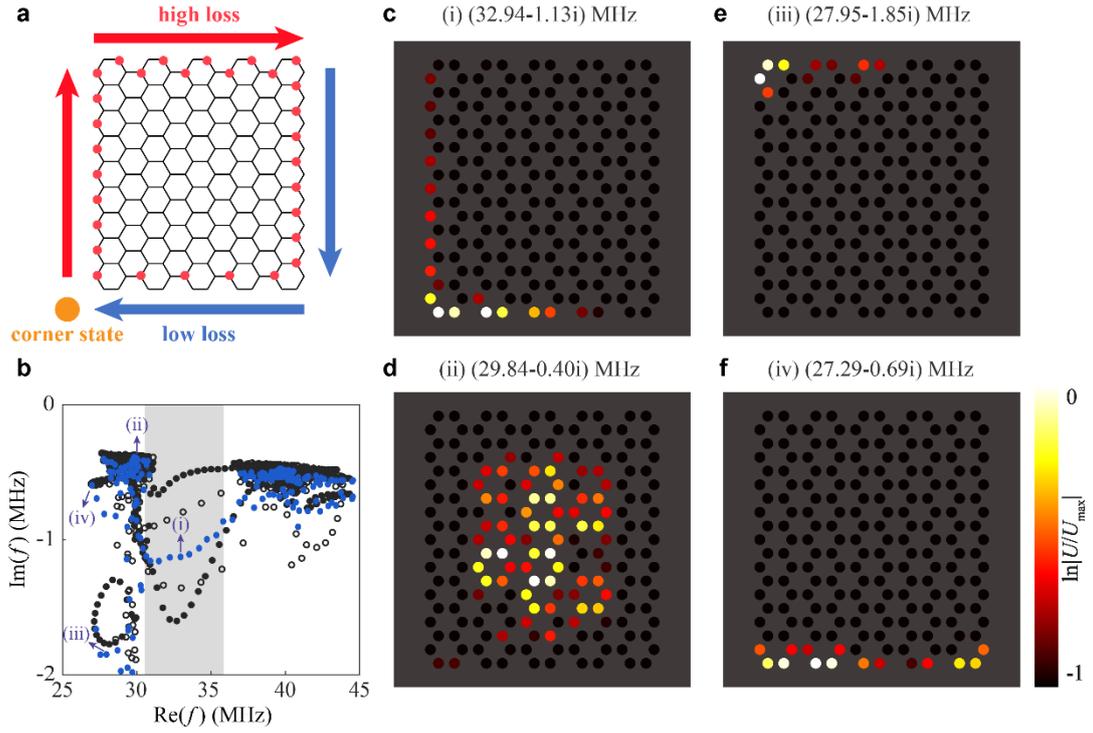

**Fig. 3| Eigenfrequencies and eigenstates of the finite-sized network. a** The network configuration. **b** The complex eigenfrequency spectra of the sample in (**a**) (blue points), the *x*-PBC/*y*-OBC supercells in Fig. 2c (black solid points) and the *x*-OBC/*y*-PBC supercells in Fig. 2e (black dashed points). **c-f** The field patterns for the eigenstates labeled from (i) to (iv) in (**b**). Patterns (i), (ii), (iii) and (iv) are corresponding to the HSTE corner state (**c**), the bulk state (**d**), the corner state originating from the non-chiral edge state at the upper edge (**e**) and the non-chiral edge state at the lower edge (**f**), respectively.

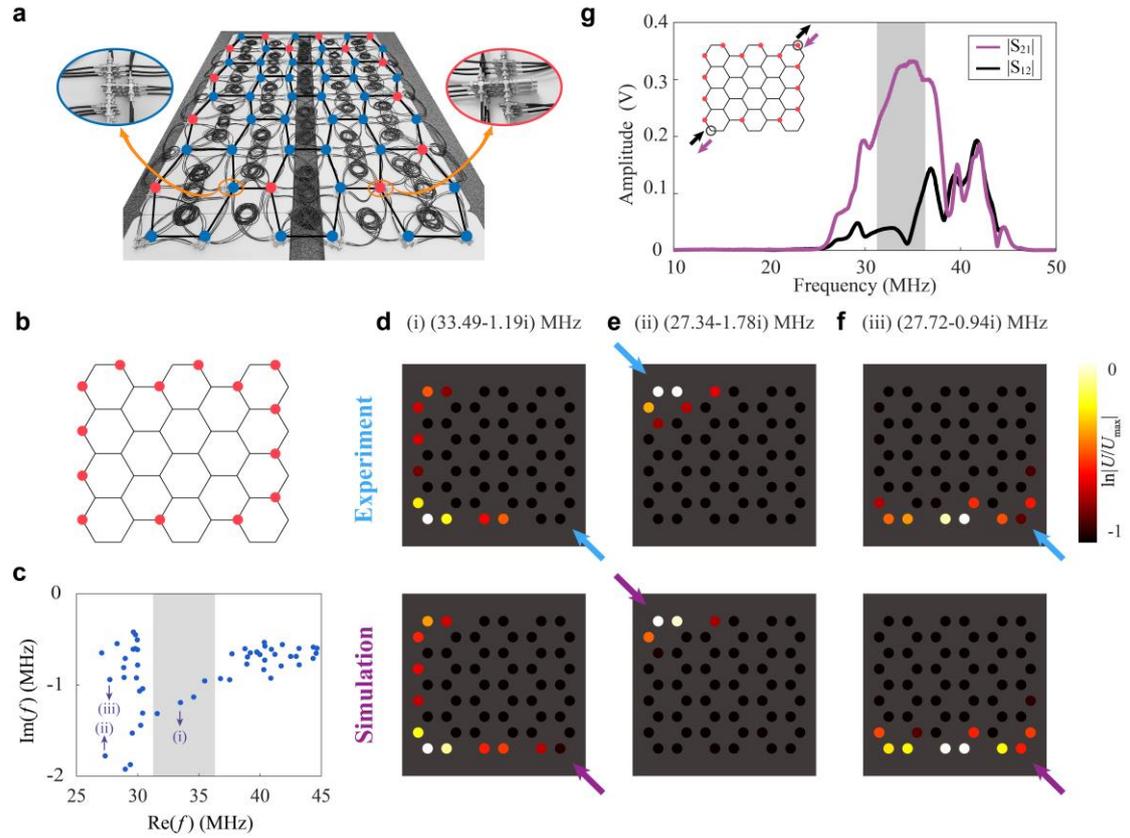

**Fig. 4| Experimental and simulated results. a** A photo of the transmission line network used in the experiment. The left and right insets show the details of the nodes without and with terminators respectively. **b** The skeleton of the experimental network. **c** The complex eigenfrequency spectrum of the sample in (**a**) and (**b**). **d-f** The experimental measured (upper panels) and numerically calculated (lower panels) field patterns of the HSTE corner state (**d**), the corner state originating from the non-chiral edge state at the upper edge (**e**), the non-chiral edge state at the lower edge (**f**), respectively. The blue and purple arrows mark the positions where the signals enter. **g** The experimental measured transmission spectra along the anti-diagonal direction.

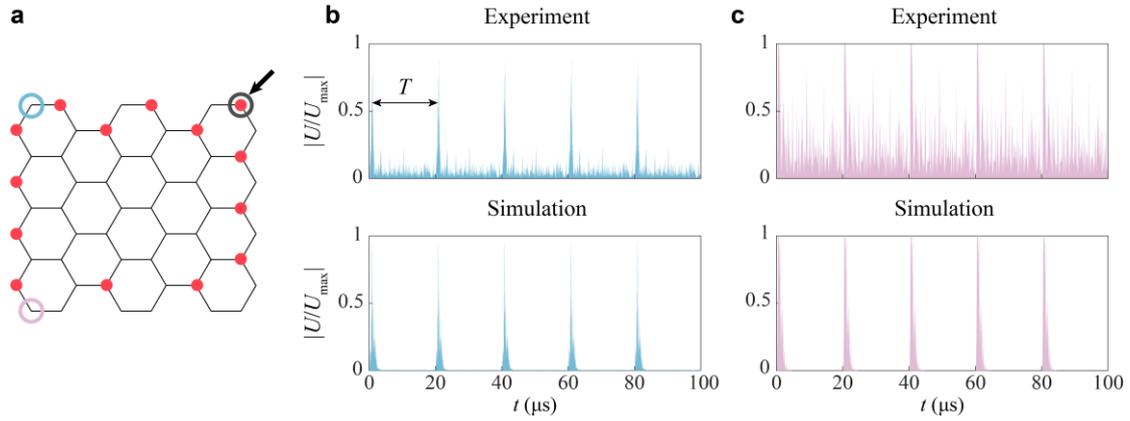

**Fig. 5| The periodicity of complex frequency synthesis. a** The locations of the excitation source and detection nodes in the network. **b-c** The time dependent response intensities $|U/U_{max}|$ at two detection nodes within 100 μs. Here, $U$ is the voltage synthesized by Eq. (13), and $U_{max}$ is the maximum synthesized voltage among all nodes.

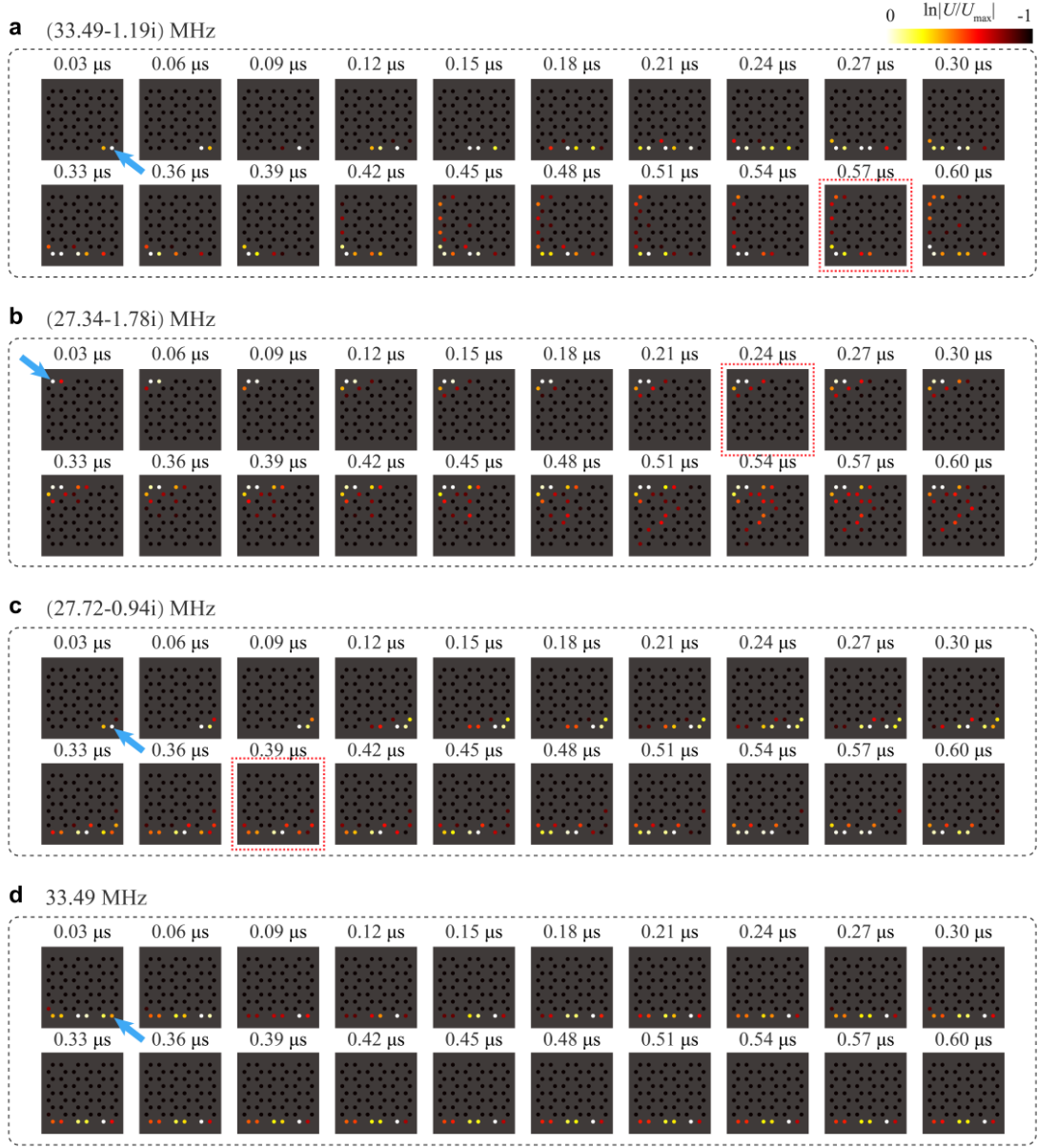

**Fig. 6| The temporal evolutions of the synthesized field patterns obtained from experimental data. a** The synthesized field patterns with frequency $f = (33.49 - 1.19i)$ MHz. The excitation points are located at the lower right corners. **b-c** The synthesized field patterns with frequencies $f = (27.34 - 1.78i)$ MHz (**b**) and $f = (27.72 - 0.94i)$ MHz (**c**), respectively. The excitation points are located at the upper left (**b**) and lower right corners (**c**), respectively. The field patterns selected for display in Fig. 4 are highlighted by the dashed red box. **d** The synthesized field patterns with a real-valued frequency $f = 33.49$ MHz.

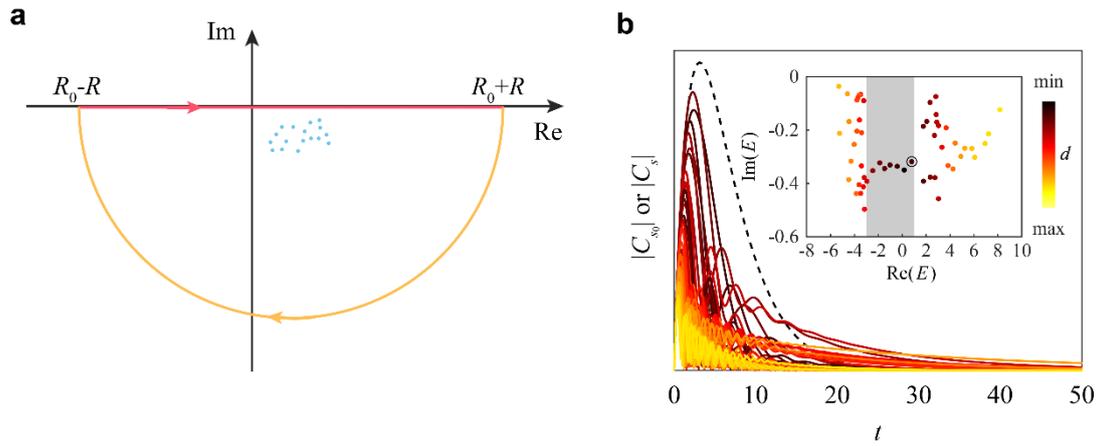

**Fig. 7| The temporal evolution of the response function. a** The schematic diagram of the loop integral in the lower half plane of the complex plane. The cyan dots represent the poles, which are situated at a considerable distance from the integral path. The red and yellow paths represent the integral paths along the real axis and the semicircular arc, respectively. **b** Temporal evolutions of coefficients $|C_{s_0}|$ and $|C_s|$ with subplot of the energy spectrum of the tight-binding model.